\documentclass[12pt,preprint]{aastex}

\usepackage{natbib}
\usepackage{graphicx}
\usepackage{amssymb}

\newcommand{\refeqn}[1]{Eq. (\ref{#1})}

\newcommand{\xt}{\tilde{x}}
\newcommand{\yt}{\tilde{y}}
\newcommand{\rt}{\tilde{r}}
\newcommand{\at}{\tilde{a}}

\newcommand{\htild}{\tilde{h}}

\newcommand{\Rt}{\tilde{R}}
\newcommand{\Apupmap}{A_{\mathrm{pupmap}}}
\newcommand{\Ain}{A_{\mathrm{in}}}
\newcommand{\Aout}{A_{\mathrm{out}}}
\newcommand{\Eout}{E_{\mathrm{out}}}
\newcommand{\Eoutk}{E_{\mathrm{out},k}}
\newcommand{\Eimg}{E_{\mathrm{img}}}
\newcommand{\Eimgk}{E_{\mathrm{img},k}}
\newcommand{\Eimgzero}{E_{\mathrm{img},0}}
\newcommand{\Eimgmk}{E_{\mathrm{img},-k}}

\newcommand{\thetat}{\tilde{\theta}}

\shorttitle{Sensitivity Analysis for Pupil Mapping}
\shortauthors{Belikov et. al.}

\begin{document}


\title{Diffraction of Apodized Pupil Mapping Systems with Aberrations}

\author{Ruslan Belikov}
\affil{Mechanical and Aerospace Engineering, Princeton University}
\email{rbelikov@princeton.edu}

\author{N. Jeremy Kasdin}
\affil{Mechanical and Aerospace Engineering, Princeton University}
\email{jkasdin@princeton.edu}

\author{Robert J. Vanderbei}
\affil{Operations Research and Financial Engineering, Princeton University}
\email{rvdb@princeton.edu}

%

\begin{abstract}
Pupil mapping is a promising and unconventional new method for high
contrast imaging  being considered for terrestrial exoplanet
searches. It employs two (or more) specially designed aspheric
mirrors to create a high-contrast amplitude profile across the telescope
pupil that does not appreciably attenuate amplitude.  As such, it
reaps significant benefits in light collecting efficiency and inner
working angle, both critical parameters for terrestrial planet
detection.  While much has been published on various aspects of
pupil mapping systems, the problem of sensitivity to wavefront
aberrations remains an open question. In this paper, we present an
efficient method for computing the sensitivity of a pupil mapped
system to Zernike aberrations. We then use this method to study the
sensitivity of a particular pupil mapping system and compare it to
the concentric-ring shaped pupil coronagraph. In particular, we show
how contrast and inner working angle degrade with increasing Zernike
order and rms amplitude, which has obvious ramifications for the
stability requirements and overall design of a planet-finding observatory.


\end{abstract}

\keywords{Extrasolar planets, coronagraphy, Fresnel propagation, diffraction
analysis, point spread function, pupil mapping, apodization, PIAA}

\section{Introduction} \label{sec:intro}

The impressive discoveries of large extrasolar planets over
the past decade has inspired widespread interest in finding
and directly imaging Earth-like planets in the habitable zones of
nearby stars.  In fact, NASA has plans to launch two space
telescopes to accomplish this,  the Terrestrial Planet Finder
Coronagraph (TPF-C) and the Terrestrial Planet Finder Interferometer
(TPF-I), while the European Space Agency is planning a similar
multi-satellite mission called Darwin. These missions are currently
in the concept study phase.  In addition, numerous ground-based
searches are proceeding using both coronagraphic and interferometric
approaches.

Direct imaging of Earth-like extrasolar planets poses extremely challenging
problems.  To see why, consider viewing our solar system from a
good distance (say from another nearby star system), our Sun would appear
$10^{10}$ times brighter than Earth.  Hence, we need an imaging system capable
of detecting planets that are $10$ orders of magnitude fainter than the star
they orbit.  Such a system is referred to as a high-contrast imager.  Furthermore, given the distances involved, the angular separation
for most targets is very small, requiring the largest launchable telescope
possible.

For TPF-C, for example, the current baseline design involves a traditional {\em
Lyot coronagraph} consisting of a modern 8th-order occulting mask attached to
the back end of a Ritchey-Chretien telescope having an $8$m by
$3.5$m elliptical primary mirror (see, e.g., \citet{KCG04}).
Alternative innovative back-end designs still being considered
include {\em shaped pupils} (see, e.g., \citet{KVSL02} and
\citet{VKS04}), a {\em visible nuller} (see, e.g., \citet{SLSWL04})
and {\em pupil mapping} (see, e.g., \citet{Guy03} where this
technique is called phase-induced amplitude apodization or PIAA). By
pupil mapping we mean a system of two lenses, or mirrors, that take
a flat input field at the entrance pupil and produce an output field
that is amplitude modified but still flat in phase (at least for
on-axis sources).

Pupil mapping has received considerable attention recently because
of its high throughput and small effective inner working angle
(IWA). These benefits could potentially permit more observations
over the mission lifetime, or conversely, a smaller and cheaper
overall telescope. As a result, there have been numerous studies
over the past year to examine the performance of pupil mapping
systems. In particular, in \citet{TV03} and \cite{VT04}, formulas for
the optical surfaces were derived using ray optics. However, this
analysis failed to account for the complete diffraction through a
pupil mapping system. More recently, \citet{Van05} provided a
detailed diffraction analysis. Unfortunately, this analysis  showed
that a pure pupil mapping system cannot achieve the needed
$10^{-10}$ contrast;  the diffraction effects from the pupil mapping
systems themselves are so detrimental that contrast attained at the
first image plane is limited to $10^{-5}$.  In \cite{GPGMRW05} and
\citet{PGRMWBG06}, a hybrid pupil mapping system was proposed that
combines the pupil mapping mirrors with a slight apodization of oversized
entrance and exit pupils. This combination does indeed achieve the
needed high-contrast point spread function (PSF). In what follows,
we will refer to such apodized pupil mapping simply as {\em pupil
mapping}, and use the term \emph{pure pupil mapping} to refer to the
original unapodized system.

What remains to be answered is how pupil mapping behaves in the
presence of optical aberrations.  It is essential that contrast be
maintained during an observation while the system is being aberrated
due to the small dynamic perturbations of the primary mirror.  An
understanding of this sensitivity is critical to the design of TPF-C
or any other observatory. In \citet{Green04}, a detailed sensitivity
analysis of both shaped pupils and Lyot coronagraphs is given.
\citet{KCG04} introduced an 8th order image plane mask for the Lyot
coronagraph that reduces the sensitivity to low-order aberrations to
the level of shaped pupils. Both of these design approaches achieve
the needed sensitivity for a realizable mission.  Up to now,
however, no comparable study has been done for pupil mapping.

In this paper, we present an efficient method for computing the
effects of wavefront aberrations on pupil mapping. We begin with a
brief review  of the design of pupil mapping systems in Section
\ref{sec:review}. We then present in Section \ref{sec:math}
a semi-analytical approach to computing the PSF of a pupil-mapping
system in the presence of aberrations represented by Zernike
polynomials. The assumption of Zernikes allows an analytical
treatment of the azimuthal angle, reducing the computational problem
from a double integral to a few single ones, thus eliminating the
need for massive computing power. We also show how these methods can
be used to compute PSFs for a purely apodizing coronagraph in
Section \ref{sec:sensApod} as well as off-axis PSFs in Section
\ref{sec:offaxis}. We conclude Section \ref{sec:huygens} with a
sensitivity analysis of pupil mapping systems in Section
\ref{sec:analysis of sensitivity}, showing the degradation of
contrast and inner working angle with Zernike order and rms
amplitude. We also demonstrate the achievable contrast limit of the
current pupil mapping approach. In Section \ref{sec:sensConc}, we
repeat the sensitivity analysis for an alternative coronagraph
design, the shaped pupil coronagraph with concentric ring masks, and
compare to pupil mapping. In particular, we show that the pupil
mapping approach to high-contrast is more sensitive to aberrations
than a purely apodized one, possibly making it more difficult to
exploit the inner working angle advantage of pupil mapping.

\section{Review of Pupil Mapping and Apodization} \label{sec:review}

\subsection{Pure Pupil Mapping via Ray Optics}
We begin by summarizing the ray-optics description of  pure pupil mapping.
An on-axis ray entering the first pupil at radius $r$ from the
center is to be mapped to radius $\rt = \Rt(r)$ at the exit pupil.
Optical elements at the two pupils ensure that the exit ray is
parallel to the entering ray. The function $\Rt(r)$ is assumed to be
positive and increasing or, sometimes, negative and decreasing.  In
either case, the function has an inverse that allows us to recapture
$r$ as a function of $\rt$: $r = R(\rt)$. The purpose of pupil
mapping is to create nontrivial amplitude profiles. An amplitude
profile function $A(\rt)$ specifies the ratio between the output
amplitude at $\rt$ to the input amplitude at $r$ (in a pure
pupil-mapping system the input amplitude is constant).
\citet{VT04} showed that for any amplitude profile $A(\rt)$ there is a
pupil mapping function $R(\rt)$ that achieves this profile.
Specifically, the pupil mapping is given by
\begin{equation}\label{1}
    R(\rt) = \pm \sqrt{\int_0^{\rt} 2 A^2(s) s ds} .
\end{equation}
Furthermore, if we consider the case of a pair of lenses that are plano on
their outward-facing surfaces, 
then the inward-facing surface profiles, $h(r)$ and $\htild(\rt)$, that
are required to obtain the desired pupil mapping are given by the solutions to
the following ordinary differential equations:
\begin{equation}\label{2}
    \frac{\partial h}{\partial r}(r)
    = \frac{r-\Rt(r)}{ \sqrt{Q_0^2 + (n^2-1)(r-\Rt(r))^2} },
    \qquad
    h(0) = z,
\end{equation}
and
\begin{equation}\label{3}
    \frac{\partial \htild}{\partial \rt}(\rt)
    = \frac{R(\rt)-\rt}{ \sqrt{\phantom{\tilde{|}}Q_0^2 + (n^2-1)(R(\rt)-\rt)^2} },
    \qquad
    \htild(0) = 0.
\end{equation}
Here, $n$ is the refractive index and
$Q_0$ is a constant determined by the distance $z$ separating the centers
($r=0$, $\rt = 0$) of the two lenses:
$Q_0 = -(n-1)z$.

Let $S(r,\rt)$ denote the distance between a point on the
first lens surface $r$ units from the center and the corresponding point on
the second lens surface $\rt$ units from its center.  Up to an additive
constant, the optical path length of a ray that exits at radius
$\rt$ after entering at radius $r = R(\rt)$ is given by
\begin{equation}\label{4}
    Q_0(\rt) = S(R(\rt),\rt) + n(\htild(\rt)-h(R(\rt))) .
\end{equation}
 \citet{VT04} showed that, for an on-axis source,
$Q_0(\rt)$ is constant and equal to $Q_0$.

\subsection{High-Contrast Amplitude Profiles} \label{sec:apod}

If we assume that a collimated beam with amplitude profile $A(\rt)$
such as one obtains as the output of a pupil mapping system
is passed into an ideal imaging system with focal length $f$, the electric
field $E(\rho)$ at the image plane is given by the Fourier transform
of $A(\rt)$:
\begin{equation}\label{5}
    E(\xi,\eta)
    =
    \frac{E_0}{\lambda f} \int\int e^{-2 \pi i \frac{\xt\xi + \yt\eta}{\lambda f}}
          A(\sqrt{\xt^2+\yt^2}) d\yt d\xt.
\end{equation}
Here, $E_0$ is the input amplitude which, unless otherwise noted,
we take to be unity.
Since the optics are azimuthally symmetric, it is convenient to use polar
coordinates.  The amplitude profile $A$ is a function of
$\rt = \sqrt{\xt^2+\yt^2}$ and the image-plane electric field depends only on
image-plane radius $\rho = \sqrt{\xi^2 + \eta^2}$:
\begin{eqnarray}
    E(\rho)
    & = &
    \frac{1}{\lambda f}
    \int\int e^{-2 \pi i \frac{\rt \rho}{\lambda f} \cos(\theta - \phi)}
                A(\rt) \rt d\theta d\rt \label{6} \\
    & = &
    \frac{2 \pi}{\lambda f}
    \int J_0\left(-2 \pi \frac{\rt \rho}{\lambda f}\right)
               A(\rt) \rt d\rt . \label{7}
\end{eqnarray}
The point-spread function (PSF) is the square of the electric field:
\begin{equation}\label{8}
    \mbox{Psf}(\rho) = |E(\rho)|^2 .
\end{equation}
For the purpose of terrestrial planet finding, it is important to construct an
amplitude profile for which the PSF at small nonzero angles is ten orders of
magnitude reduced from its value at zero.
 \citet{VSK03} explains how these functions are computed as
solutions to certain optimization problems.

\begin{figure}
\begin{center}
\mbox{\includegraphics[width=3.0in]{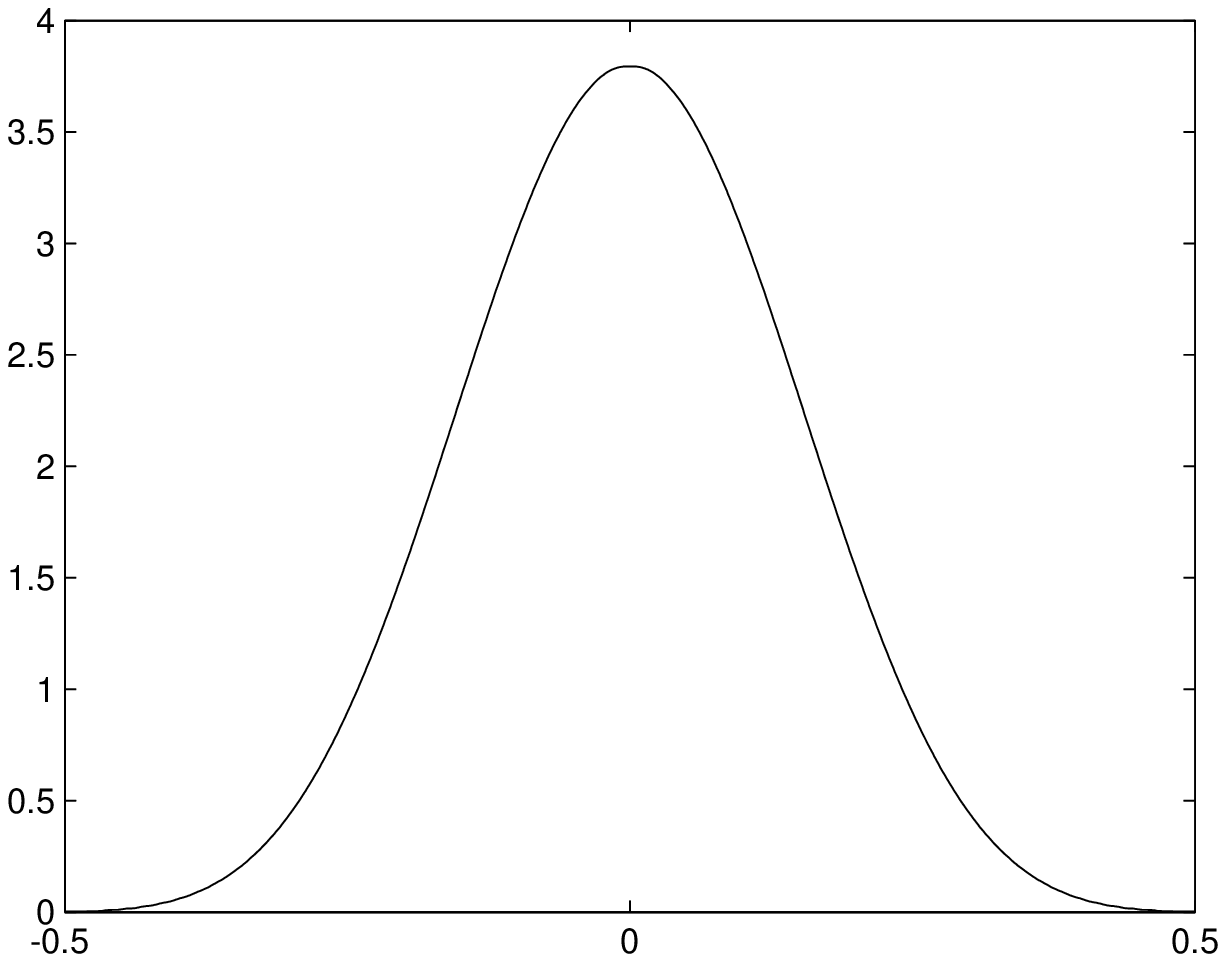}}
\mbox{\includegraphics[width=3.0in]{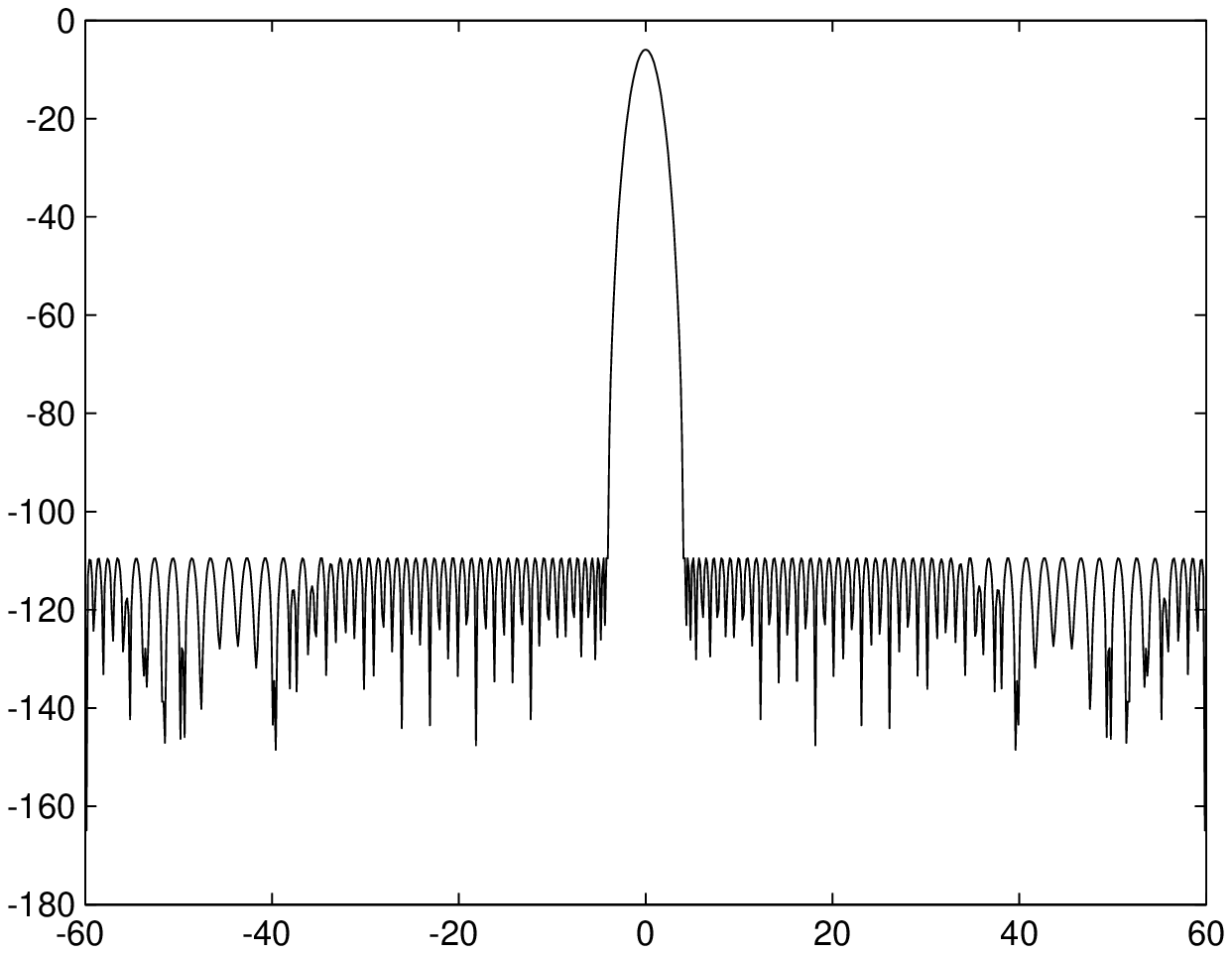}}
\end{center}
\caption{{\em Left.}
An amplitude profile providing contrast of $10^{-10}$ at tight inner
working angles.
{\em Right.} The corresponding on-axis point spread function.
}
\label{fig:1}
\end{figure}

The high-contrast amplitude profile used in the rest of this paper is shown in
Figure \ref{fig:1}.

We end this section by noting that the need to design customized
amplitude profiles arise in many areas, usually in the context of
apodization---i.e., profiles that only attenuate the beam.
\citet{ref:Slepian} was perhaps the first to study this problem
carefully. For some recent applications, the reader is refered to
the following papers in the area of beam shaping: \citet{CG99},
\citet{GOPP02} and \citet{HJ03}.

\subsection{Apodized Pupil Mapping Systems} \label{sec:hybrid}

\citet{Van05} showed that pure pupil mapping systems designed for
contrast of $10^{-10}$ actually achieve much less than this due to
harmful diffraction effects that are not captured by the simple ray
tracing analysis outlined in the previous section.  For most systems
of practical real-world interest (i.e., systems with apertures of a
few inches and designed for visible light), contrast is limited to
about $10^{-5}$.  In \citet{Van05}, certain hybrid designs were
considered that improve on this level of performance but none of the
hybrid designs presented there completely overcame this
diffraction-induced contrast degradation.

In this section, we describe an apodized pupil mapping system that
is somewhat more complicated than the designs presented in
\citet{Van05}. This hybrid design, based on ideas proposed by
Olivier Guyon and Eugene Pluzhnik (see \citet{PGRMWBG06}), involves
three additional components.  They are
\begin{enumerate}
    \item a preapodizer $A_0$
    to soften the edge of the first lens/mirror so as to
    minimize diffraction effects caused by hard edges,
    \item a postapodizer to smooth out low spatial frequency ripples produced
    by diffraction effects induced by the pupil mapping system itself, and
    \item a backend phase shifter to smooth out low spatial frequency ripples
    in phase.
\end{enumerate}
Note that the backend phase shifter can be built into the second lens/mirror.
There are several choices for the preapodizer.
For this paper, we use the preapodizer given by Eqs. (3) and (4) in
\citet{PGRMWBG06}:
\[
    A_0(r) = \frac{A(r)(1 + \beta)}{A(r) + \beta A_{\mathrm{max}}},
\]
where $A_{\mathrm{max}}$ denotes the maximum value of $A(r)$ and $\beta$ is a
scalar parameter, which we take to be $0.1$.  It is easy to see that
\begin{itemize}
    \item $A(r)/A_{\mathrm{max}} \le A_0(r) \le 1$ for all $r$,
    \item $A_0(r)$ approaches $1$ as $A(r)$ approaches $A_{\mathrm{max}}$, and
    \item $A_0(r)$ approaches $0$ as $A(r)$ approaches $0$.
\end{itemize}

Incorporating a post-apodizer introduces a degree of freedom
that is lacking in a pure pupil mapping system.  Namely, it is possible to
design the pupil mapping system based on an arbitrary amplitude profile and
then convert this profile to a high-contrast profile via an
appropriate choice of backend
apodizer.  We have found that a simple Gaussian amplitude profile that
approximately matches a high-contrast profile works very well.  Specifically,
we used
\[
    \Apupmap(\rt) = 3.35 e^{-22(\rt/\at)^2},
\]
where $\at$ denotes the radius of the second lens/mirror.

The backend apodization is computed by taking the actual output
amplitude profile as computed by a careful diffraction analysis,
smoothing it by convolution with a Gaussian distribution, and then
apodizing according to the ratio of the desired high-contrast
amplitude profile $A(\rt)$ divided by
the smoothed output profile.  Of course, since a true apodization can never
intensify a beam, this ratio must be further scaled down so that it is nowhere
greater than unity.  The Gaussian convolution kernel we used has mean zero and
standard deviation $\at/\sqrt{100,000}$.

The backend phase modification is computed by a similar smoothing operation
applied to the output phase profile.  Of course, the smoothed output phase
profile (measured in radians) must be converted to a surface profile
(having units of length).  This conversion requires us
to assume a certain specific wavelength.  As a consequence, the resulting
design is correct only at one wavelength.  The ability of the system
to achieve high contrast
degrades as one moves away from the design wavelength.

\begin{figure}
\begin{center}
\mbox{\includegraphics[width=6.0in]{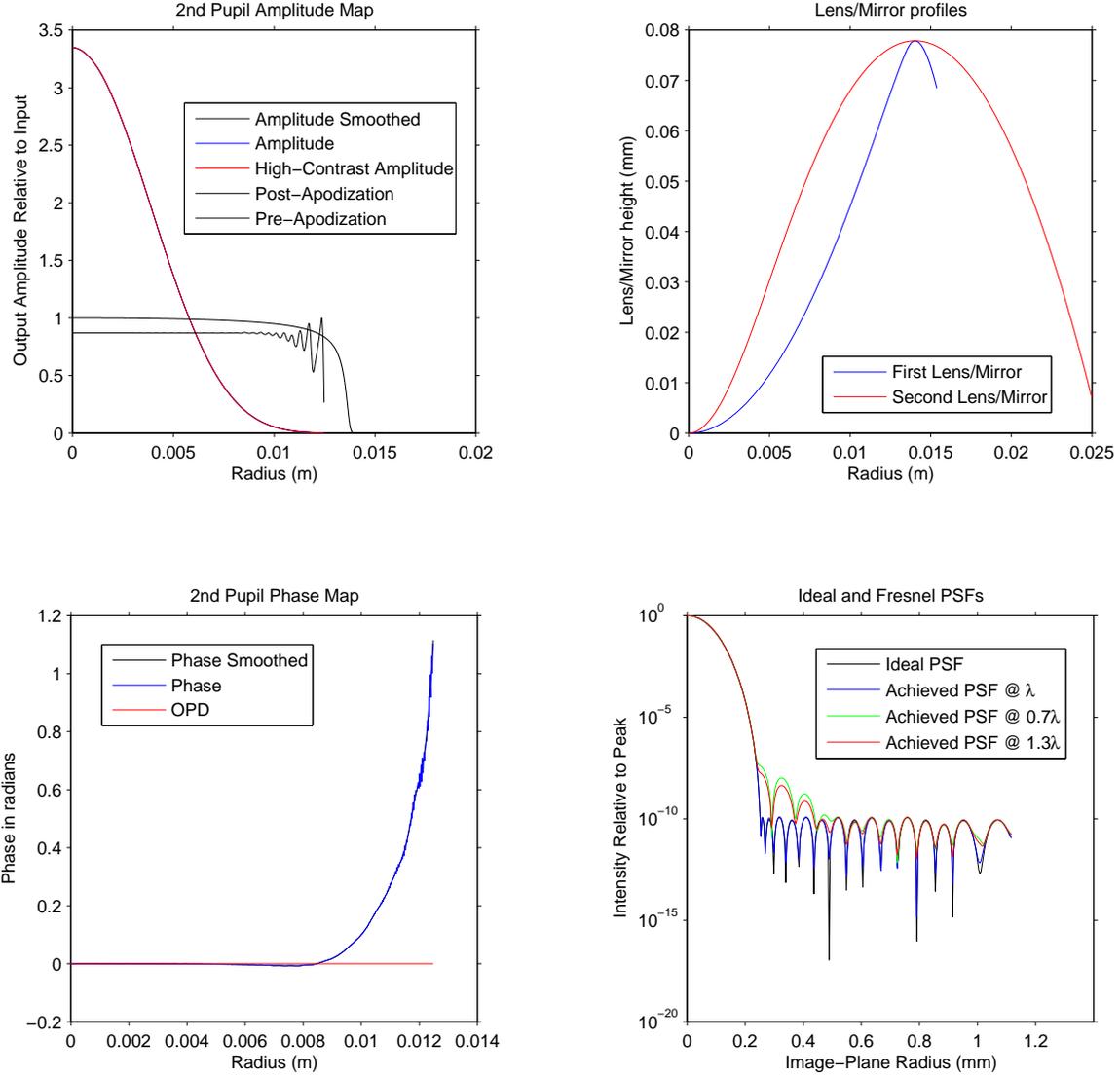}}
~
\end{center}
\caption{ Hybrid pupil mapping system as described in Section
\ref{sec:hybrid}. {\em Upper-left} plot shows in red the target
high-contrast amplitude profile and in blue the Gaussian profile
that is a close approximation to it. The two black plots show the
pre- and post-apodizers. {\em Upper-right} plot shows the lens
profiles, blue for the first lens and red for the second. {\em
Lower-left} plot shows in blue the phase map computed using Huygens
propagation and in black the smoothed approximation. {\em
Lower-right} plot shows in black the ideal PSF, in blue the actual
PSF at the design wavelength, in red the actual PSF at $130\%$ of
the design wavelength, and in green the actual PSF at $70\%$ of the
design wavelength. (The horizontal scale for $130\%$ and $70\%$ has
been dilated by factors of 1/1.3 and 1/0.7, respectively, and the
x-axis labels correspond to the original $\lambda$.)} \label{fig:2}
\end{figure}

Figure \ref{fig:2} shows plots characterizing the performance of the apodized
pupil mapping system described in this section.
The specifications for this system are as follows.
The designed-for wavelength is $632.8$nm.
The optical elements are assumed to be mirrors separated by $0.375$m.
The system is an on-axis system and we therefore make the non-physical
assumption that the mirrors don't obstruct the beam.
That is, the mirrors are invisible except when they are needed.
The mirrors take as input a $0.025$m on-axis beam and produce a $0.025$m
pupil-remapped exit beam.
The second mirror is oversized by a factor of two;
that is, its diameter is $0.050$m.
The postapodizer ensures that only the central half contributes to
the exit beam.
The first mirror is also oversized appropriately as shown in
the upper-right subplot of Figure \ref{fig:2}.

After the second mirror, the exit beam is brought to a focus. The focal length is $2.5$m. The
lower-right subplot in Figure \ref{fig:2} shows the ideal PSF (in black)
together with the achieved PSF at three wavelengths: at $70\%$
(green), $100\%$ (blue), and $130\%$ (red) of the design wavelength.
At the design wavelength, the achieved PSF matches the ideal PSF
almost exactly. Note that there is minor degradation at the other
two wavelengths mostly at low spatial frequencies.

It is important to note that the PSFs in Figure \ref{fig:2} correspond to a bright
on-axis source (i.e., a star).  Off-axis sources, such as faint
planets, undergo two effects in a pupil mapping system that differ from the response of
a conventional imaging system: an effective magnification and a
distortion. These are explained in detail in \citet{VT04} and
\citet{TV03}. The magnification, in particular, is due to an effective
narrowing of the exit pupil as compared to the entrance pupil.
It is this magnification that provides pupil mapped systems their smaller effective inner working angle.
The techniques in Section \ref{sec:huygens} will allow us to
compute the exact off-axis diffraction pattern of an apodized pupil mapped coronagraph and thus to see these effects.


While the effective magnification of a pupil mapping system results in
an inner working angle advantage of about a factor of two,  it does not produce high-quaity  diffraction limited images of
off-axis sources because of the distortion inherent in the system. \citet{Guy03} proposed the following solution to
this problem. He suggested using this system merely as a mechanism
for concentrating (on-axis) starlight in an image plane. He then
proposed that an occulter be placed in the image plane to remove the starlight. All
other light, such as the distorted off-axis planet light, would be allowed
to pass through the image plane. On the back side would be a second, identical
pupil mapping system  (with the
apodizers removed), that would ``umap'' the off-axis beam and thus remove the
distortions introduced by the first system (except for some beam walk---see \citet{VT04}).

\section{Diffraction Analysis of Apodized Pupil Mapping with an Aberrated Wavefront} \label{sec:huygens}

A general 2D diffraction analysis of a pupil mapping system with a
possibly aberrated input field requires significant computing power.
In this section, we present a much more efficient method assuming
the aberrations are Zernike polynomials, which allows us to treat the
azimuthal variables analytically. In particular, we perform a full
end-to-end diffraction analysis of an apodized pupil mapping system
with aberrations, as well as with a tilted field such as one from an
off-axis planet. Such an analysis is required to determine the
sensitivity of these systems to phase errors at the input pupil, as
well as the off-axis response. At the end of this section, we
describe the results for specific cases of interest.

\subsection{Mathematical Development} \label{sec:math}

If we assume that a flat, on-axis, electric field arrives at the entrance
pupil, then the electric field at a particular point of the exit pupil
can be well-approximated by superimposing the phase-shifted waves from each
point across the entrance pupil (this is the well-known Huygens-Fresnel
principle---see, e.g., Section 8.2 in \cite{BW99}).  For an apodized pupil
mapping system, we can write this as
\begin{equation}\label{9}
  \Eout(\xt, \yt)
  =
  \Aout(\rt)
  \int\int
       \frac{1}{\lambda Q(\xt,\yt,x,y)}
       e^{ 2 \pi i Q(\xt,\yt,x,y) /\lambda }
       \Ain(r)
       dydx,
\end{equation}
where
\begin{equation}\label{20}
       Q(\xt,\yt,x,y) =
       \sqrt{(x-\xt)^2+(y-\yt)^2+(h(r)-\htild(\rt))^2}
       + n(Z - h(r) + \htild(\rt))
\end{equation}
is the optical path length, $Z$ is the distance between the plano lens
surfaces (i.e., a constant slightly larger than $z$), $\Ain(r)$ denotes the
input amplitude apodization at radius $r$, $\Aout(\rt)$ denotes the output
amplitude apodization at radius $\rt$,
and where, of course, we have used $r$ and $\rt$ as shorthands for the radii in
the entrance and exit pupils, respectively.
If the arriving field is not flat but instead has a phase profile given by
$\alpha(x,y)$, then the integral must include this phase shift:
\begin{equation}\label{50}
  \Eout(\xt, \yt)
  =
  \Aout(\rt)
  \int\int
       \frac{1}{\lambda Q(\xt,\yt,x,y)}
       e^{ 2 \pi i Q(\xt,\yt,x,y)/\lambda + i\alpha(x,y)}
       \Ain(r)
       dydx.
\end{equation}
As before, it is convenient to work in polar coordinates:
\begin{equation}\label{10}
  \Eout(\rt,\thetat)
  =
  \Aout(\rt)
  \int\int
    \frac{1}{\lambda Q(\rt,r,\theta-\thetat)}
        e^{ 2 \pi i Q(\rt,r,\theta-\thetat)/\lambda + i\alpha(r,\theta)}
        \Ain(r)
    r d\theta dr,
\end{equation}
where
\begin{equation}\label{21}
    Q(\rt,r,\theta) =
    \sqrt{r^2-2r\rt\cos\theta+\rt^2+(h(r)-\htild(\rt))^2}
        + n(Z - h(r) + \htild(\rt)) .
\end{equation}
For numerical tractability, it is essential to make approximations so that
the integral over $\theta$ can be carried out analytically.
To this end, we need to make an appropriate approximation to the square root
term:
\begin{equation}\label{13}
     S = \sqrt{r^2-2r\rt\cos\theta+\rt^2+(h(r)-\htild(\rt))^2} .
\end{equation}

We approximate the $1/Q(\rt,r,\theta-\thetat)$
amplitude-reduction
factor in \refeqn{10} by the constant $1/Z$ (the {\em paraxial approximation}).
The $Q(\rt,r,\theta-\thetat)$ appearing in the exponential must,
on the other hand, be treated with care.
Recall that $Q(\rt,R(\rt),0)$ is a constant.  Since constant phase shifts are
immaterial, we can subtract it from $Q(\rt,r,\theta)$ in \refeqn{10} to get
\begin{equation} \label{140}
  \Eout(\rt,\thetat)
  \approx
  \frac{\Aout(\rt)}{\lambda Z}
  \int\int
       e^{ 2 \pi i \left(Q(\rt,r,\theta-\thetat)-Q(\rt,R(\rt),0)\right)/\lambda
            + i\alpha(r,\theta)}
        \Ain(r)
    r d\theta dr.
\end{equation}
Next, we write the difference in $Q$'s as follows:
\begin{eqnarray}
    Q(\rt,r,\theta-\thetat)-Q(\rt,R(\rt),0)
    & = &
    S(\rt,r,\theta-\thetat)-S(\rt,R(\rt),0)
    + n(h(R(\rt)) - h(r)) \label{141} \nonumber \\
    & = &
    \frac{S^2(\rt,r,\theta-\thetat)
              -S^2(\rt,R(\rt),0)}{S(\rt,r,\theta-\thetat)+S(\rt,R(\rt),0)}
    + n(h(R(\rt)) - h(r)) \label{142}
\end{eqnarray}
and then we expand out the numerator and cancel big terms that can be
subtracted one from another to get
\begin{eqnarray}
    S^2(\rt,r,\theta-\thetat)-S^2(\rt,R(\rt),0)
    & = &
    (r - R(\rt)) (r + R(\rt)) - 2\rt\left(r \cos (\theta-\thetat)
      - R(\rt)\right)
    \nonumber\\
    && \quad +
    \left(h(r)-h(R(\rt))\right) \left(h(r)+h(R(\rt))-2\htild(\rt)\right) .
\end{eqnarray}
When $r = R(\rt)$ and $\theta = 0$,
the right-hand side clearly vanishes as it should.
Furthermore, for $r$ close to $R(\rt)$ and $\theta$ close to $\thetat$,
the right-hand side gives an accurate formula for computing the deviation from
zero.  That is to say, the right-hand side is easy to program in such a manner
as to avoid subtracting one large number from another, which is always the
biggest danger in numerical computation.

So far, everything is exact (except for the paraxial approximation).
The only further approximation we make is to replace
$S(\rt,r,\theta-\thetat)$
in the denominator of \refeqn{142} with $S(\rt,R(\rt),0)$
so that the denominator becomes just $2S(\rt,R(\rt),0)$.
Since $S(\rt,R(\rt),0)$ appears many times in coming formulas, we abbreviate
it as $S(\rt)$ and hope this does not create any confusion.
Putting this altogether, we get a new approximation, which we refer to as
the {\em Huygens} approximation:
\begin{eqnarray}
  \Eout(\rt,\thetat)
  & \approx &
  \frac{\Aout(\rt)}{\lambda Z}
  \int\int
    e^{ 2 \pi i \left(
      \frac{
        (r - R(\rt)) (r + R(\rt)) + 2\rt R(\rt)
        +
        \left(h(r)-h(R(\rt))\right) \left(h(r)+h(R(\rt))-2\htild(\rt)\right)
      }{2S(\rt)}
      + n(h(R(\rt)) - h(r))
    \right) /\lambda} \nonumber \\
    && \qquad \qquad \times
    e^{2 \pi i \left( -\frac{\rt r \cos(\theta - \thetat)}{S(\rt)}
                      \right) /\lambda + i\alpha(r,\theta)}
    \Ain(r)
    d\theta r dr \label{144} \\
    & = &
    \frac{\Aout(\rt)}{\lambda Z}
    \int K(r,\rt) L(r,\rt,\thetat) \Ain(r) r dr , \label{160}
\end{eqnarray}
where
\begin{eqnarray}
    K(r,\rt)
    & = &
    e^{ 2 \pi i \left(
      \frac{
        (r - R(\rt)) (r + R(\rt)) + 2\rt R(\rt)
        +
        \left(h(r)-h(R(\rt))\right) \left(h(r)+h(R(\rt))-2\htild(\rt)\right)
      }{2S(\rt)}
      + n(h(R(\rt)) - h(r))
    \right) /\lambda} \nonumber \\
    L(r,\rt,\thetat)
    & = &
    \int_0^{2 \pi}
    e^{2 \pi i \left( -\frac{\rt r \cos(\theta - \thetat)}{S(\rt)}
                      \right) /\lambda + i\alpha(r,\thetat)}
    d\theta. \label{145}
\end{eqnarray}

Suppose now that the aberration is given by the $(l,m)$-th Zernike
polynomial
\begin{equation} \label{alpha}
    \alpha(r,\theta) = \epsilon Z_l^m(r/a) \cos(m \theta) ,
\end{equation}
where $\epsilon$ is a small number. In this case, the kernel
$L(r,\rt,\thetat)$ can be expressed in terms of Bessel functions.
Recall that the $n$-th Bessel function $J_n(x)$ is defined by the
requirement that $i^n J_n(x)$ be the $n$-th Fourier coefficient in
the Fourier series expansion of $e^{i x \cos(\theta)}$.  That is,
\begin{equation} \label{148}
    J_n(x) = \frac{1}{2 \pi i^n}
              \int_0^{2 \pi} e^{i x \cos \theta} e^{i n \theta} d \theta .
\end{equation}
The Fourier series then, of course, is simply
\begin{equation} \label{149}
    e^{i x \cos \theta} = \sum_{k= -\infty}^{\infty} i^k J_k(x) e^{i k \theta}.
\end{equation}
This series is usually referred to as the {\em Jacobi-Anger expansion}.

Plugging the Zernike polynomial into the definition of the kernel
$L(r,\rt,\thetat)$ and substituting a Jacobi-Anger expansion for each of the two
cosine-exponentials, we get that
\begin{eqnarray}
    L(r,\rt,\thetat)
    &=&
    \int_0^{2 \pi}
    e^{2 \pi i \left( -\frac{\rt r \cos(\theta - \thetat)}{S(\rt)}
                      \right) /\lambda + i\alpha(r,\theta)}
    d\theta \label{146} \\
    &=&
    \sum_j \sum_k \int i^j i^k
    J_j\left(-\frac{2\pi \rt r}{\lambda S(\rt)}\right)
    e^{ij(\theta-\thetat)}
    J_k\left( \epsilon Z_l^m(r/a)\right)
    e^{ikm\theta}
    d \theta \nonumber \\
    &=&
    \sum_j \sum_k
    i^{j+k}
    J_j\left(-\frac{2\pi \rt r}{\lambda S(\rt)}\right)
    J_k\left( \epsilon Z_l^m(r/a)\right)
    e^{-ij\thetat}
    \int
    e^{ij\theta+ikm\theta}
    d \theta
    \nonumber \\
    &=&
    2 \pi
    \sum_k
    i^{k(1-m)}
    J_{km}\left(\frac{2\pi \rt r}{\lambda S(\rt)}\right)
    J_k\left( \epsilon Z_l^m(r/a)\right)
    e^{ikm\thetat}
    \nonumber \\
    &=&
    \sum_k
    e^{ikm\thetat}
    L_k(r,\rt)
    \label{147} ,
\end{eqnarray}
where
\begin{equation}
    L_k(r,\rt)
    =
    2 \pi
    i^{k(1-m)}
    J_{km}\left(\frac{2\pi \rt r}{\lambda S(\rt)}\right)
    J_k\left( \epsilon Z_l^m(r/a)\right)
\end{equation}
The second to the last equality above follows from the trivial fact that
\[
    \int_0^{2 \pi} e^{in \theta} d\theta
    =
    \left\{ \begin{array}{ll}
        2 \pi & \quad n = 0 \\
    0     & \quad \mbox{otherwise} .
    \end{array}\right.
\]
Substituting \refeqn{147} into \refeqn{160}, we can write
\begin{eqnarray}
    \Eout(\rt,\thetat) &=& \frac{\Aout(\rt)}{\lambda Z}
                   \sum_k e^{ikm\thetat}
                   \int K(r,\rt)L_k(r,\rt) \Ain(r) r dr \nonumber \\
                       &=& \sum_k e^{ikm\thetat} \Eoutk(\rt),
\end{eqnarray}
where
\begin{equation}
    \Eoutk(\rt)
    =
    \frac{\Aout(\rt)}{\lambda Z} \int K(r,\rt)L_k(r,\rt) \Ain(r) r dr .
\end{equation}

Ignoring a complex unit exponential, the electric field in the image plane is
just the Fourier transform of the electric field in the pupil plane.  Working
in polar coordinates, we have
\begin{eqnarray}
    \Eimg(\rho,\phi)
    & = &
    \frac{1}{\lambda f}
    \int\int
    e^{\frac{-2 \pi i}{\lambda f} \rt \rho \cos(\thetat - \phi)}
    \Eout(\rt,\thetat) d\thetat \rt d\rt
    \nonumber \\
    & = &
    \frac{1}{\lambda f}
    \sum_k
    \int\int
    e^{\frac{-2 \pi i}{\lambda f} \rt \rho \cos(\thetat - \phi)}
    e^{ikm\thetat}
    \Eoutk(\rt)
    d\thetat \rt d\rt \label{161} .
\end{eqnarray}
From the definition (\refeqn{148})
of the Bessel functions as certain Fourier coefficients, we get
\begin{eqnarray}
    \int
    e^{ikm\thetat}
        e^{\frac{-2 \pi i}{\lambda f} \rt \rho \cos(\thetat - \phi)}
    d\thetat
    &=&
    \int
    e^{ikm(\thetat+\phi)}
        e^{\frac{-2 \pi i}{\lambda f} \rt \rho \cos(\thetat)}
    d\thetat
    \nonumber \\
    &=&
    2 \pi i^{km} e^{ikm\phi}
    J_{km}\left( -\frac{2 \pi}{\lambda f} \rt \rho \right) . \label{150}
\end{eqnarray}
Substituting \refeqn{150} into \refeqn{161}, we get
\begin{eqnarray}
    \Eimg(\rho,\phi)
    & = &
    \frac{2 \pi}{\lambda f}
    \sum_k
    e^{ikm\phi}
    \int
    i^{km}
    J_{km}\left( -\frac{2 \pi}{\lambda f} \rt \rho \right)  \nonumber
    \Eoutk(\rt)
    \rt d\rt \label{152} , \\
    & = &
    \sum_k
    e^{ikm\phi}
    \Eimgk(\rho) , \label{154}
\end{eqnarray}
where
\begin{eqnarray}
    \Eimgk(\rho)
    :=
    \frac{2 \pi}{\lambda f}
    \int
    i^{km}
    J_{km}\left( -\frac{2 \pi}{\lambda f} \rt \rho \right)
    \Eoutk(\rt)
    \rt d\rt \label{153} .
\end{eqnarray}
Finally, since $\Eimgk(\rho)$ = $\Eimgmk(\rho)$, we see that
\begin{eqnarray}
    \Eimg(\rho,\phi)
    & = &
    \Eimgzero(\rho)
    +
    2
    \sum_{k=1}^{\infty}
    \cos(km\phi)
    \Eimgk(\rho)
    . \label{155}
\end{eqnarray}

Note that
\[
    |J_k(x)| \approx \frac{1}{k!} \left(\frac{x}{2}\right)^k
\]
for $0 \le x \ll 1$.  Hence, if we assume that $\epsilon  < 10^{-3}$,
then the error incurred by dropping all terms in the sum
on $k$ for which $k>2$ will be approximately $10^{-6}$ at the most.  Hence,
any error in the PSF will be at the $10^{-12}$ level.

Thus, for a given Zernike polynomial aberration at the entrance pupil and
$\epsilon$ in \refeqn{alpha}, \refeqn{155} can be used to compute
the resulting response of the apodized pupil mapping system.
Note that an arbitrary aberration can be expressed as a linear
combination of Zernike polynomials and, for each Zernike, the
azimuthal variable can be integrated explicitly, thereby reducing 2D
integrals to sums of 1D integrals.

\subsection{Pure Apodization as a Special Case of Apodized Pupil Mapping} \label{sec:sensApod}

If we set $\Apupmap(\rt) = 1$ for all $\rt$, then the pupil
mapping system reduces to a trivial forward propagation of a flat
beam.  That is, the lenses/mirrors have perfectly flat surfaces and
no remapping occurs.  Hence, in this case, the hybrid system becomes
a pure apodization system with the desired high-contrast apodization
being achieved by a pair of apodizers, one appearing before the beam
is propagated and one after (except for the Fresnel diffraction due
to the propagation between the two mirrors.) In this fashion, the
mathematics developed in \ref{sec:math} can be applied to a pure
pupil-apodization coronagraph.

\subsection{Off-Axis Diffraction Response} \label{sec:offaxis}

\begin{figure}
\begin{center}
\mbox{\includegraphics[width=3.0in]{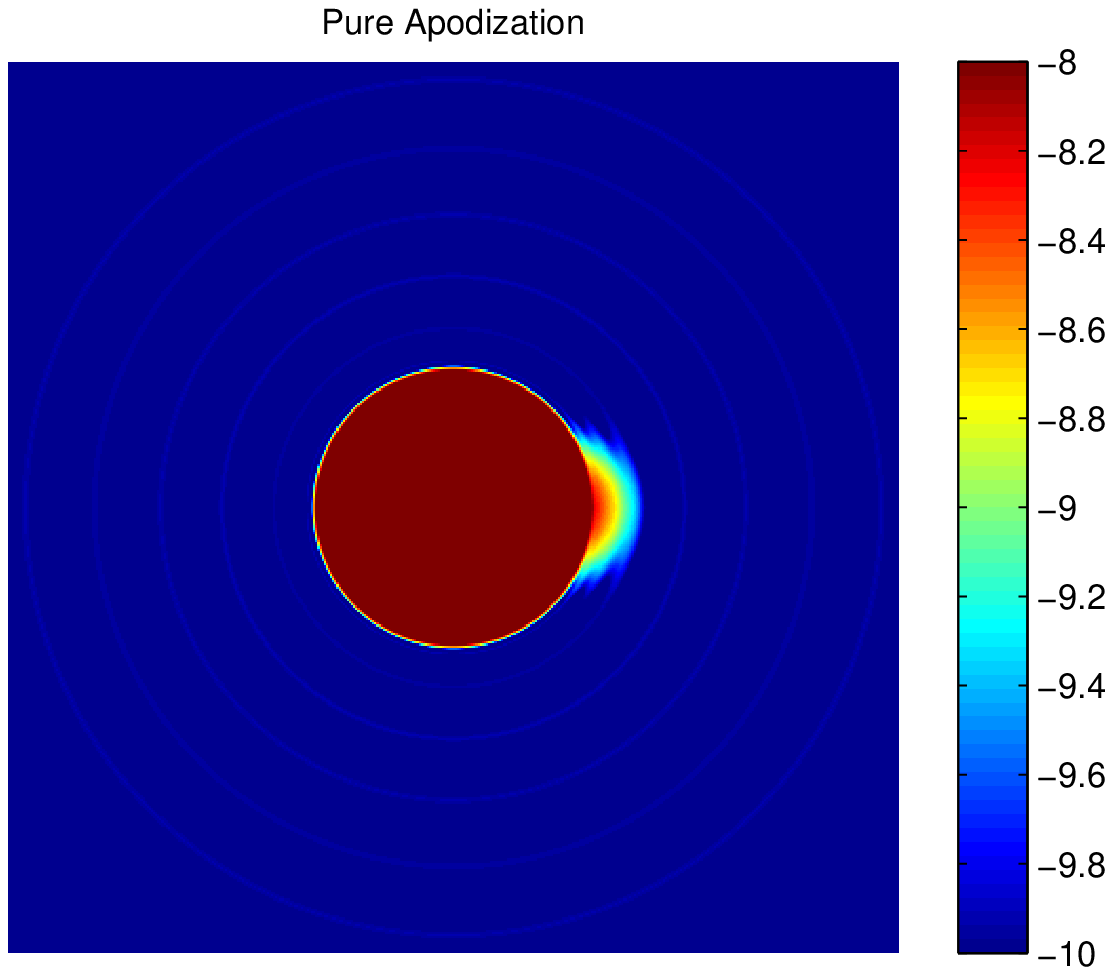}}
\mbox{\includegraphics[width=3.0in]{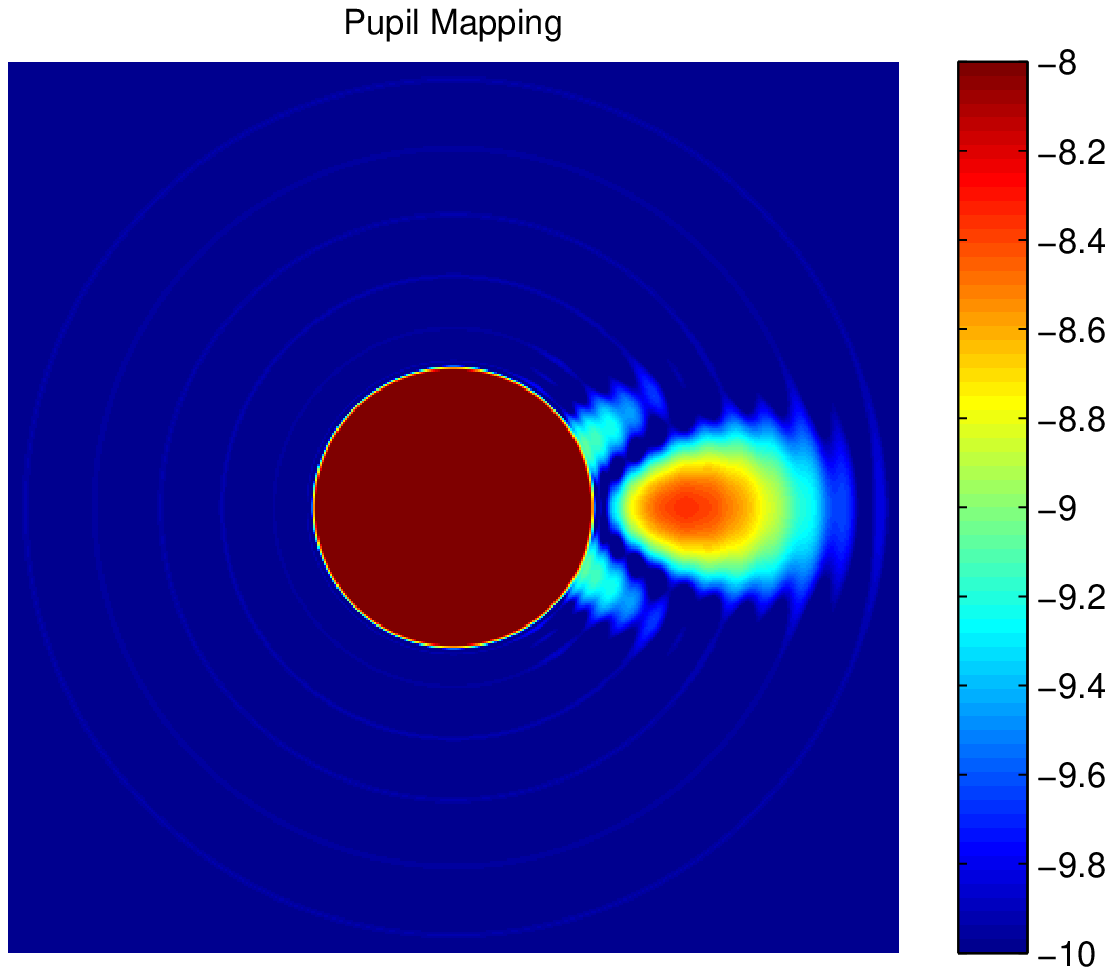}}
\mbox{\includegraphics[width=4.0in]{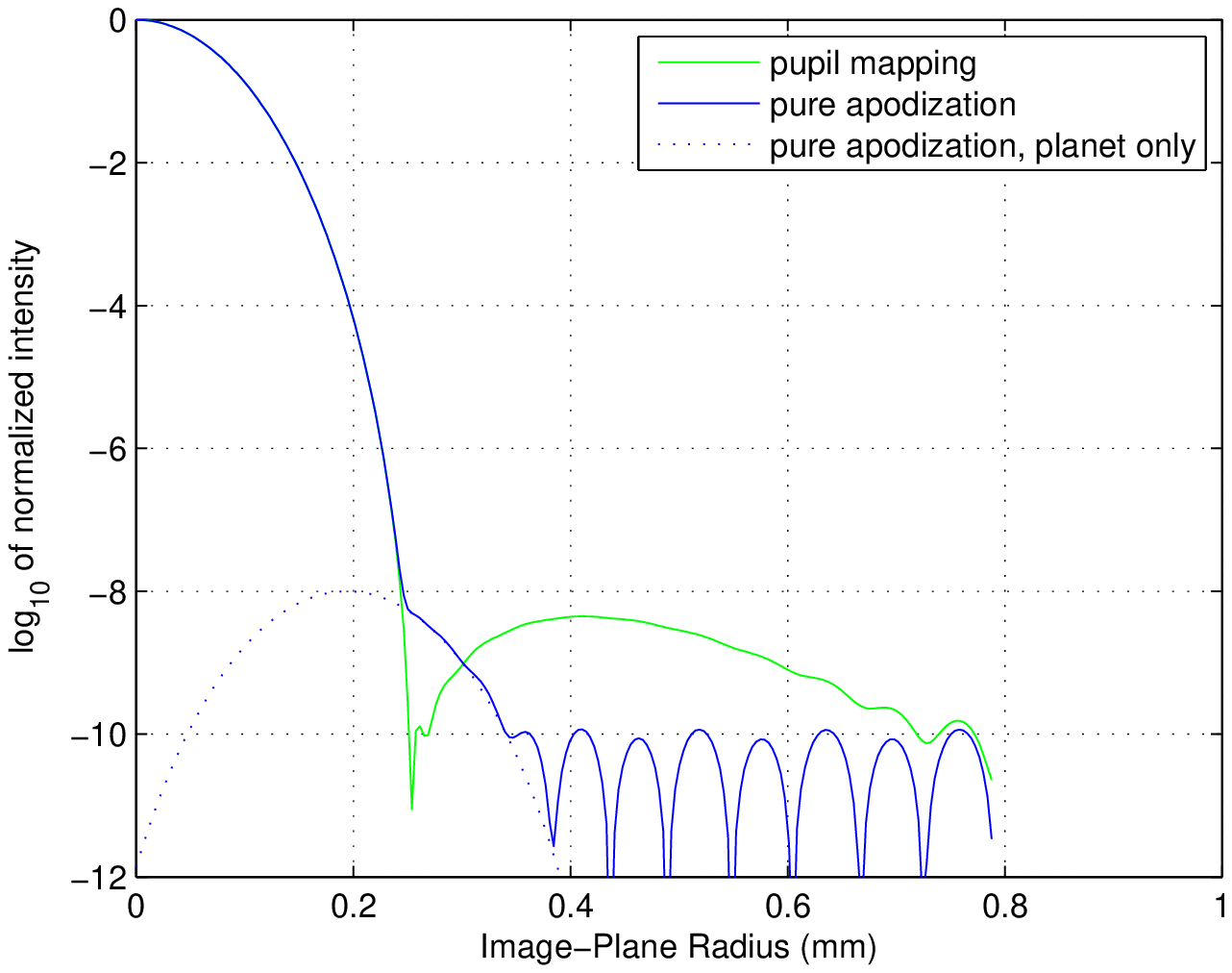}}
~
\end{center}
\caption{Distortion of off-axis sources in pupil mapping. {\em Top}:
images of a $3\lambda/D$ planet (e.g. a 1au planet viewed from 10
parsecs away with a 4m telescope). The star appears as a saturated
disk. Both images are contrast plots normalized to the peak
intensity of the star image. The z scale is $log_{10}$ of contrast.
{\em Bottom}: Horizontal slices. Note that these are contrast plots,
i.e. everything is normalized to the peak intensity of the star
image in pupil mapping as well as pure apodization. On an absolute
intensity scale, the whole pupil mapping curve would be higher due
to better throughput.} \label{fig:tilt}
\end{figure}

In addition to modeling optical aberrations, the equations in
Section \ref{sec:math} can be used to model off-axis sources such as
planets because they are mathematically equivalent to large tilt
aberrations. For example, a 1au planet at 10pc is equivalent to a
tilt aberration of $0.1$ arcseconds. For a $D = 4$ meter aperture and
$\lambda = 632.8$nm, this would correspond to an angle of $3
\lambda/D$ and $\epsilon = 9.63$ in \refeqn{alpha}. Figure
\ref{fig:tilt} is a contrast plot showing the result of computing
\refeqn{155} for such a planet, along with a simulation for a pure
pupil-apodizing coronagraph. In addition, an on-axis star was added
in both cases that is $10^8$ times brighter than the planet. We
assumed a coronagraph entrance aperture of $25$mm, and a focal length
of $2.5$m, so that for the apodizing coronagraph, or any conventional
imaging system, the planet appears at roughly $194\mu$m off-axis in
the image plane.

As mentioned before, pupil mapping both magnifies, which is good, and
distorts, which is bad. Both effects are evident in Figure \ref{fig:tilt}. The
magnification moves the center of the distorted planet image out by
roughly a factor of $2$ (to about 400$\mu$m in Figure
\ref{fig:tilt}).
Thus, the  inner working angle of the system is
effectively reduced, which is one of the benefits of pupil
mapping. On the other hand, pupil mapping distorts the planet,
which reduces the peak intensity a little with respect to the star's peak
intensity. However, note that Figure \ref{fig:tilt} shows contrast
plots, i.e. intensity is normalized to peak star intensity for each
case. On an absolute intensity scale, the whole pupil mapping curve
will be higher because of the throughput advantage. Also note that
the distortion caused by the two-mirror system can be countered by
inverting the pupil mapping after starlight has been blocked, as
described at the end of Section \ref{sec:hybrid}.

\subsection{Analysis of Sensitivity to Zernike Aberrations}
\label{sec:analysis of sensitivity}

Figures \ref{fig:3} to \ref{fig:6} show sensitivity results for the
apodized pupil mapping system described in Section \ref{sec:hybrid}.
Figure \ref{fig:3} is for aberrations that are $1/100$-th wave rms
whereas Figure \ref{fig:4} is for smaller $1/1000$-th wave rms
aberrations.  Figure \ref{fig:5} shows radial cross-sections and
Figure \ref{fig:6} shows differences between the radial cross-sections of
the aberrated and un-aberrated PSF's. (As noted, the $(0,0)$, or
piston, Zernike is differenced with the ideal PSF.) Note that these
plots show the PSF for an on-axis source, i.e. a star. By comparing
these plots to Figure \ref{fig:tilt}, one can directly compare the
levels of the aberrations to the PSF of an off-axis planet.

Two important effects are evident in Figure \ref{fig:6}.  First, the
$(0,0)$, or piston, Zernike represents the nominally achieved
contrast and so its difference with the ideal shows the limitation
of the current design approach. Any additional errors due to the
Zernike aberrations, as shown in the subsequent subplots (which are
differenced with the nominal $(0,0)$  PSF), add to this static limit.
Second, the pupil-mapping system shows a degradation in inner
working angle with increasing Zernike and increasing aberration
size. Any  design that attempts to  exploit the inherent
magnification to achieve a smaller inner working angle must be able
to achieve a higher level of aberration control.

Another important issue to consider is the sensitivity to
aberrations at wavelengths other than the design wavelength.
Comparing Figure \ref{fig:5} with the bottom-right subplot of
Figure \ref{fig:2}, it is
evident that the degradation due to aberrations at $1/1000$ wave
level are generally higher than the degradation due to detuning of
the wavelength by a factor of $1.3$ or $0.7$. Thus, the sensitivity to
aberrations at these wavelengths should look very similar to Figure
\ref{fig:5}. Our simulations do indeed confirm this. Of course, at
the longer wavelength, the aberrated PSF in Figure \ref{fig:5}
would be expanded
by the appropriate amount, and contracted for the shorter
wavelength.

\section{Sensitivity Analysis for Concentric Ring Masks} \label{sec:sensConc}

For comparison purposes, we now consider a high-contrast imaging system based
on a simple concentric rings shaped pupil as described in \citet{VSK02}.
A concentric-ring pupil-plane mask can be thought of as a circularly symmetric
binary apodization.
Hence, we assume that we are given such an apodization function $A(r)$.

As before, we assume that the input electric field could involve some
aberration given by $\alpha(r,\theta)$.
The image plane electric field is given by the Fourier transform of the
pupil-plane field, which in polar coordinates is
\begin{equation}
    E_{\mathrm{img}}(\rho,\phi)
    =
    \frac{1}{\lambda f}
    \int\int
    e^{-\frac{2\pi i}{\lambda f}r \rho \cos(\theta-\phi)}
    e^{i\alpha(r,\theta)}
    A(r) d\theta r dr .
\end{equation}
Expanding the two exponentials in a Jacobi-Anger expansion, we get that
\begin{equation}
    E_{\mathrm{img}}(\rho,\phi)
    =
    \frac{1}{\lambda f}
    \int_0^a
    \sum_j \sum_k i^{j+k}
    J_j \left(-\frac{2\pi }{\lambda f}r \rho \right)
    J_k \left( \epsilon Z_l^m(r/a) \right)
    e^{-ij\phi}
    \left[ \int_0^{2\pi} e^{ij\theta + km\theta} d\theta \right]
    A(r) r dr  .
\end{equation}
The integral in brackets is easily solved, as before,
eliminating one of the summations.  With some rearranging, this yields
\begin{equation}
    E_{\mathrm{img}}(\rho,\phi)
    =
    \frac{2\pi}{\lambda f}
    \sum_k i^{(1-m)k}
    e^{i km \phi}
    \int_0^a
    J_{km} \left(\frac{2\pi }{\lambda f}r \rho \right)
    J_k \left( \epsilon Z_l^m(r/a) \right)
    A(r) r dr .
\label{eq:eq6}
\end{equation}
Finally, we use the symmetry property of the Bessel Functions
($J_{-m}(x) = (-1)^m J_m(x)$) to yield the image plane field,
\begin{equation}
    \Eimg(\rho,\phi)
    =
    \Eimgzero(\rho)
    +
    2
    \sum_{k=1}^{\infty}
    \cos(km\phi)
    \Eimgk(\rho)
    ,
\end{equation}
where
\begin{equation}
    \Eimgk(\rho)
    =
    \frac{2\pi}{\lambda f}
    \int_{0}^{a}
    i^{(1-m)k}
    J_{km}  \left(\frac{2\pi}{\lambda f} r \rho \right)
    J_k \left( \epsilon Z_l^m(r/a) \right)
    A(r) r dr .
\end{equation}
For a concentric ring mask having its edges at $r_j$,
the formula for $\Eimgk(\rho)$ simplifies to
\begin{equation}
    \Eimgk(\rho)
    =
    \frac{2\pi}{\lambda f}
    \sum_j \int_{r_{2j}}^{r_{2j+1}}
    i^{(1-m)k}
    J_{km}  \left(\frac{2\pi}{\lambda f} r \rho \right)
    J_k \left( \epsilon Z_l^m(r/a) \right)
    r dr .
\end{equation}

Figures \ref{fig:7} to \ref{fig:10} show sensitivity results for a
high-contrast concentric-ring pupil mask. We also computed
sensitivity results for the pure apodizing system; they appeared
virtually identical to the results shown in these figures, which
agree with the results in \citet{Green04}. Likewise, the off-axis
results are indistinguishable from those of the pure apodizing
system in Figure \ref{fig:tilt}.



\section{Conclusions} \label{conc}

\begin{figure}
\begin{center}
\mbox{\includegraphics[width=6in]{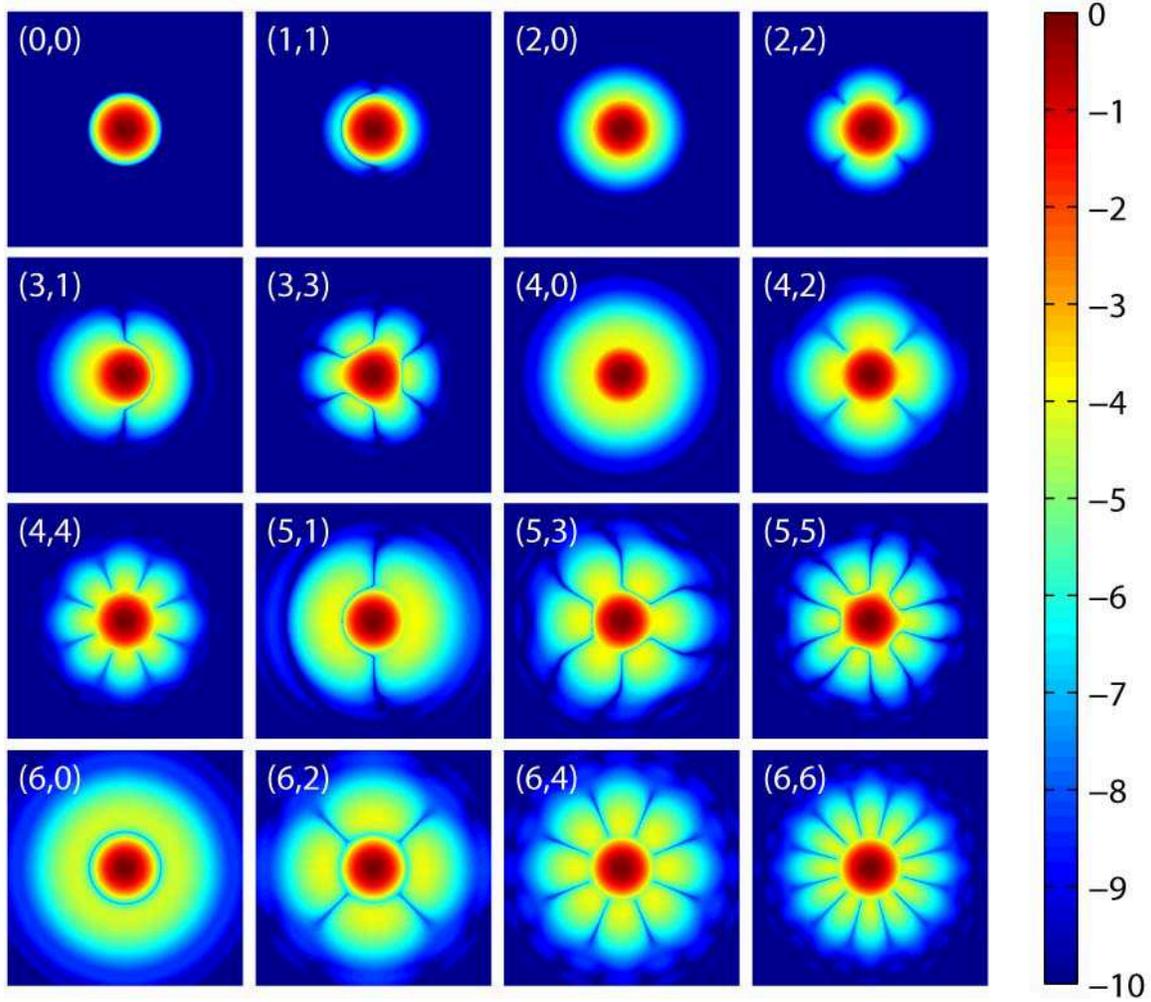}}
\end{center}
\caption{Pupil mapping. PSFs associated with various Zernikes
normalized so that the rms error is, in each case, $1/100$-th wave.
The plots correspond to: Piston $(0,0)$, Tilt $(1,1)$, Defocus
$(2,0)$, Astigmatism $(2,2)$ Coma $(3,1)$, Trefoil $(3,3)$,
Spherical Aberration $(4,0)$, Astigmatism 2nd Order $(4,2)$,
Tetrafoil $(4,4)$, etc. {\em Important note:} As explained in
Section \ref{sec:hybrid}, off-axis sources are ``pushed out'' by a
magnification factor of about $2$ in these high-contrast
pupil-mapping systems. This needs to be remembered when comparing
with later plots. } \label{fig:3}
\end{figure}

\begin{figure}
\begin{center}
\mbox{\includegraphics[width=6in]{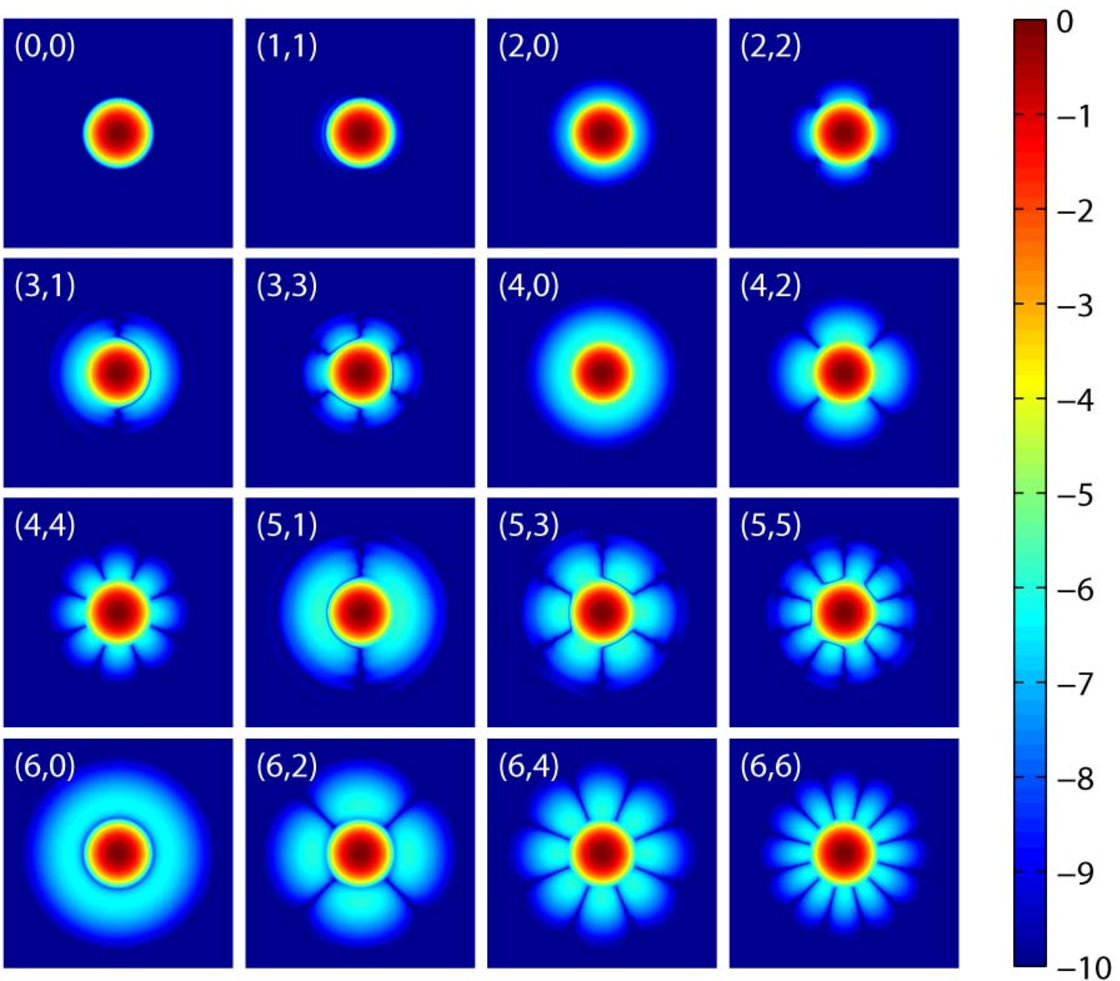}}
\end{center}
\caption{Pupil mapping. PSFs associated with various Zernikes
normalized so that the rms error is, in each case, $1/1000$-th wave.
The plots correspond to: Piston $(0,0)$, Tilt $(1,1)$, Defocus
$(2,0)$, Astigmatism $(2,2)$, Coma $(3,1)$, Trefoil $(3,3)$,
Spherical Aberration $(4,0)$, Astigmatism 2nd Order $(4,2)$,
Tetrafoil $(4,4)$, etc. } \label{fig:4}
\end{figure}

\begin{figure}
\begin{center}
\mbox{\includegraphics[width=6in]{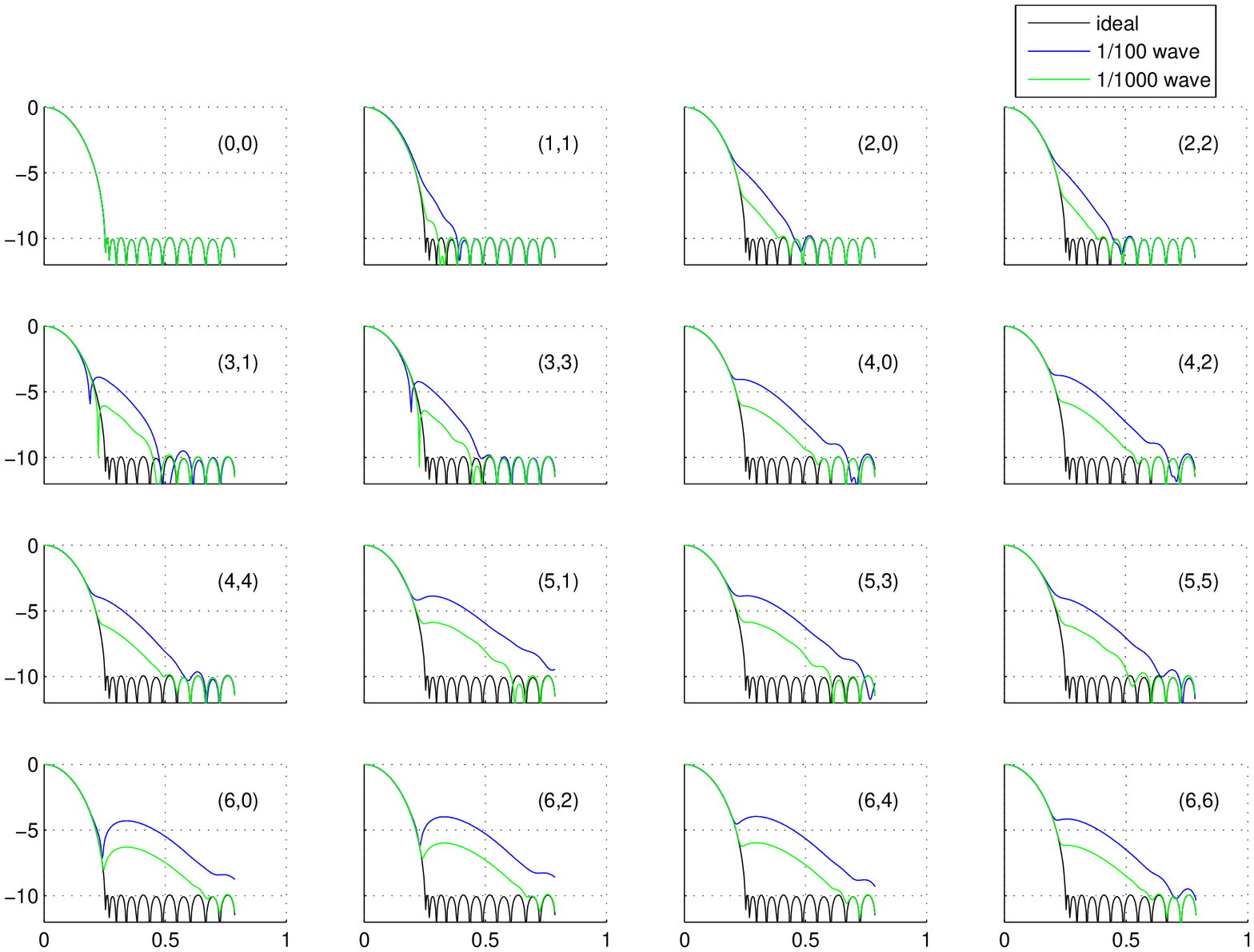}}
\end{center}
\caption{Cross-sectional 1-D plots through the positive $x$-axis in the plots
of the previous two figures.
}
\label{fig:5}
\end{figure}

\begin{figure}
\begin{center}
\mbox{\includegraphics[width=6in]{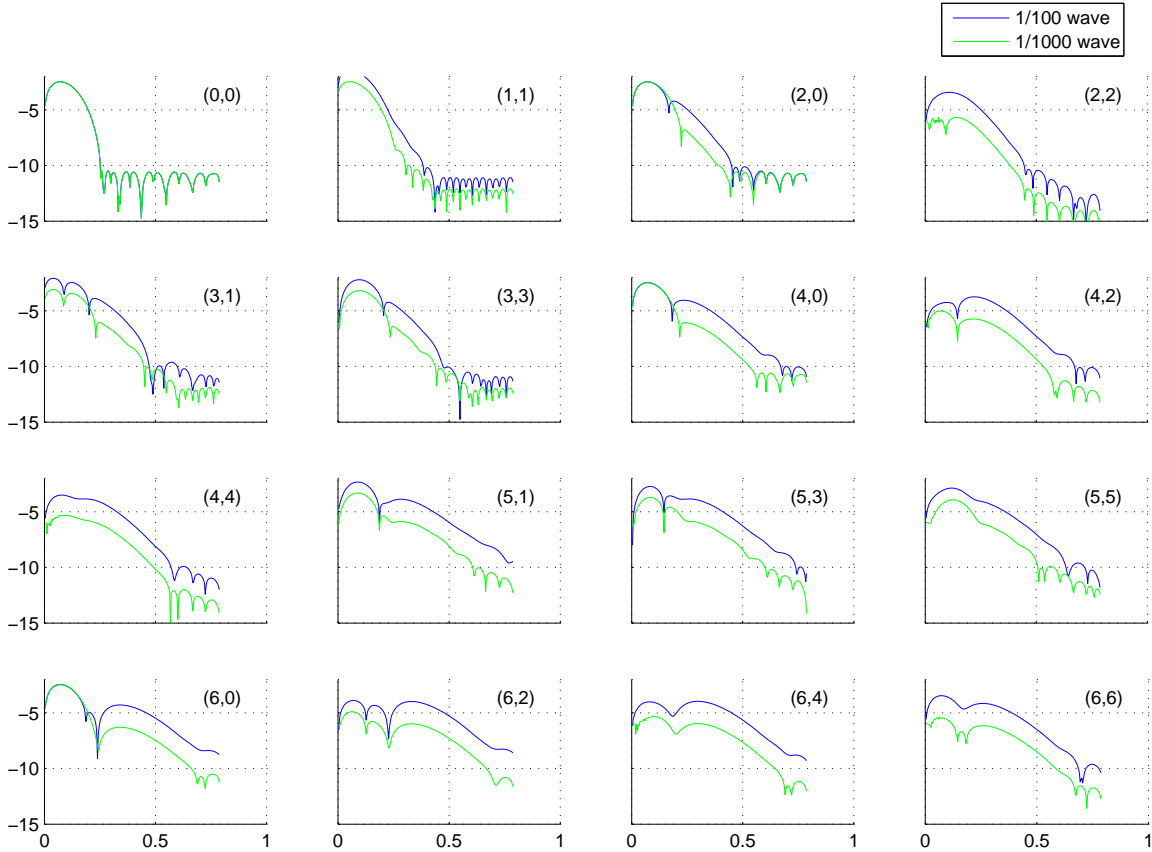}}
\end{center}
\caption{Difference plots. For all the plots except (0,0), the
difference is between the aberrated and non-aberrated profile for
pupil mapping. (The non-aberrated profile is essentially the (0,0),
or piston term). For the (0,0) difference plot in the upper left,
the difference is between the non-aberrated pupil mapping profile
and the non-aberrated ideal apodization profile, showing that pupil
mapping does not match the ideal apodization profile at the contrast
level of $10^{-11}$ and below.} \label{fig:6}
\end{figure}

\begin{figure}
\begin{center}
\mbox{\includegraphics[width=6in]{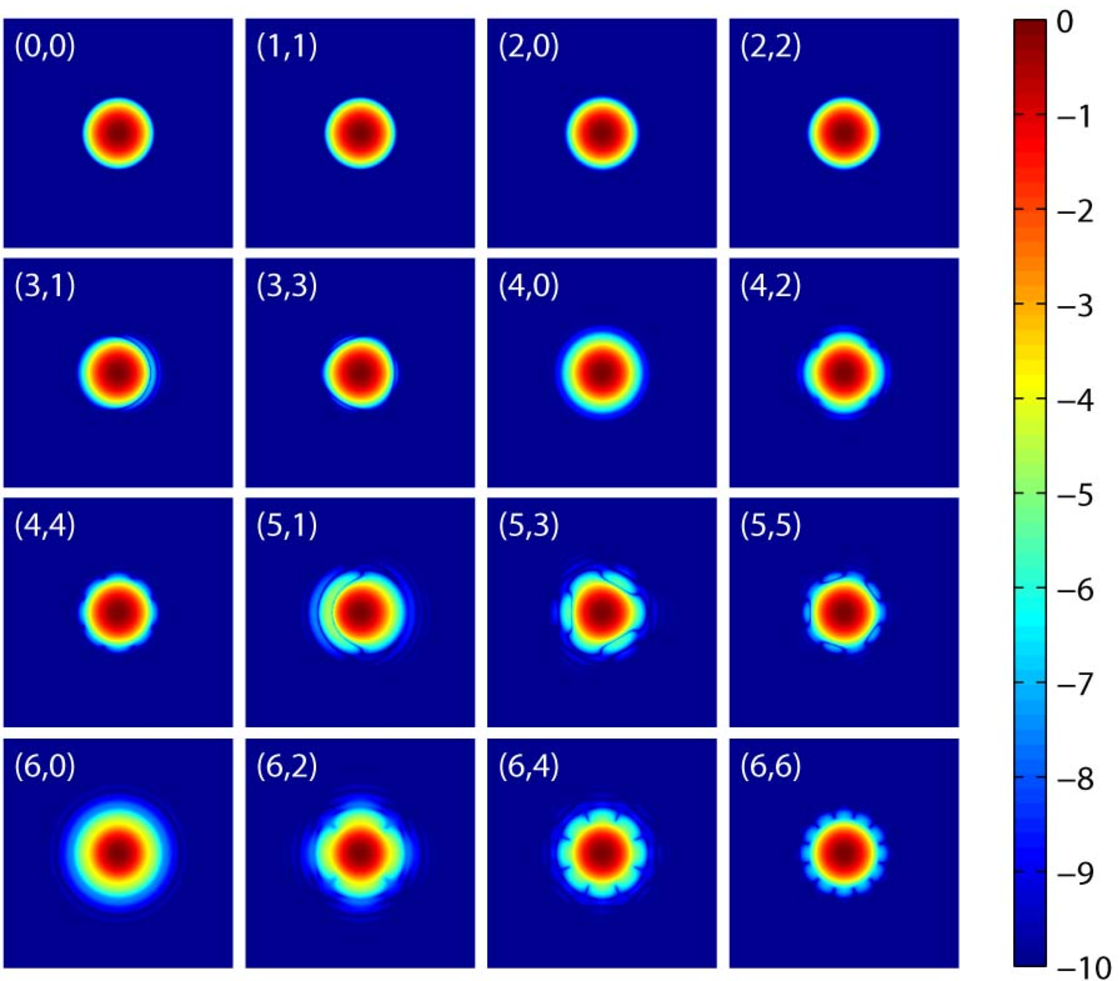}}
\end{center}
\caption{Concentric Rings. PSFs associated with various Zernikes
normalized so that the rms error is, in each case, $1/100$-th wave.
The plots correspond to: Piston $(0,0)$, Tilt $(1,1)$, Defocus
$(2,0)$, Astigmatism $(2,2)$ Coma $(3,1)$, Trefoil $(3,3)$,
Spherical Aberration $(4,0)$, Astigmatism 2nd Order $(4,2)$,
Tetrafoil $(4,4)$, etc.} \label{fig:7}
\end{figure}

\begin{figure}
\begin{center}
\mbox{\includegraphics[width=6in]{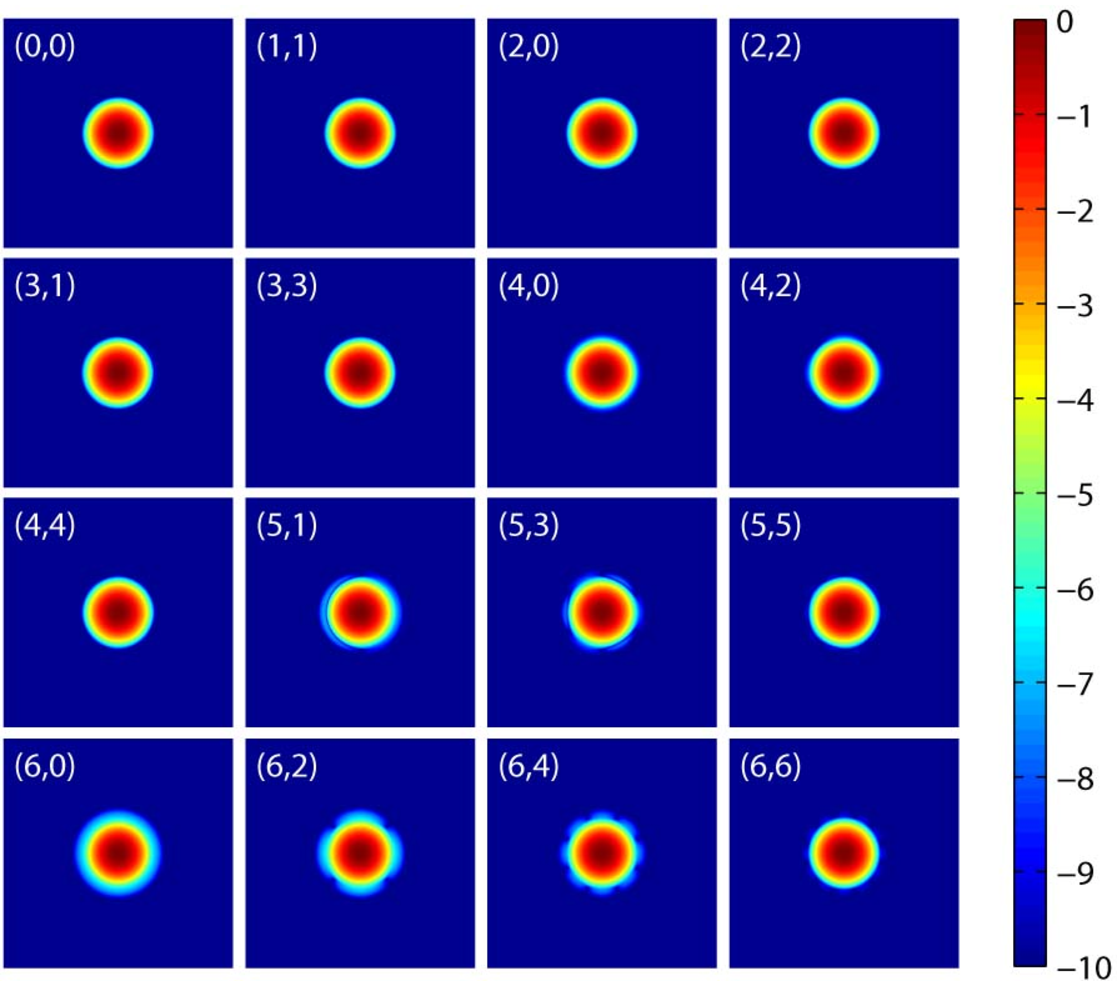}}
\end{center}
\caption{Concentric Rings. PSFs associated with various Zernikes
normalized so that the rms error is, in each case, $1/1000$-th wave.
The plots correspond to: Piston $(0,0)$, Tilt $(1,1)$, Defocus
$(2,0)$, Astigmatism $(2,2)$ Coma $(3,1)$, Trefoil $(3,3)$,
Spherical Aberration $(4,0)$, Astigmatism 2nd Order $(4,2)$,
Tetrafoil $(4,4)$, etc. } \label{fig:8}
\end{figure}

\begin{figure}
\begin{center}
\mbox{\includegraphics[width=6in]{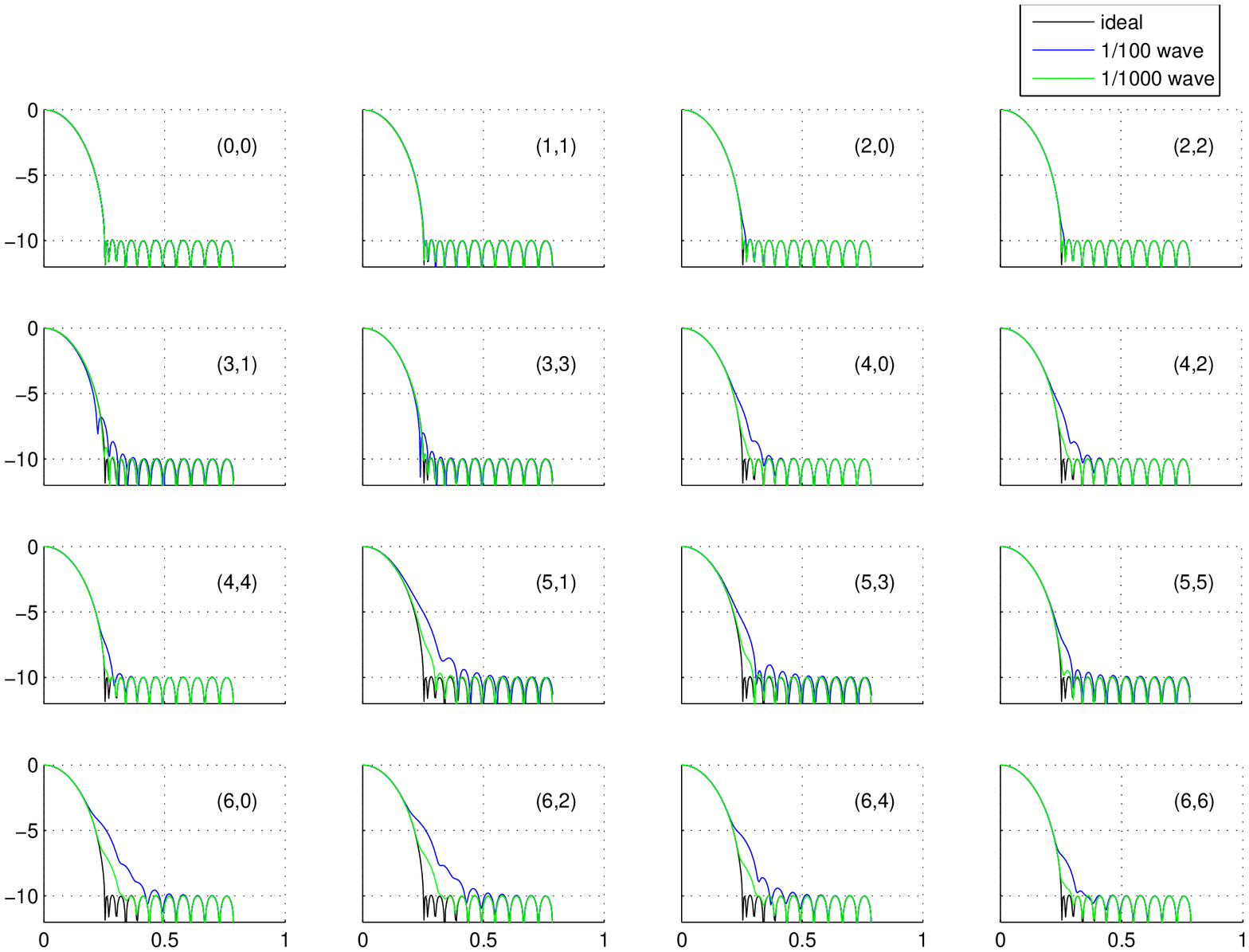}}
\end{center}
\caption{Cross-sectional 1-D plots through the positive $x$-axis in the plots
of the previous two figures.
}
\label{fig:9}
\end{figure}

\begin{figure}
\begin{center}
\mbox{\includegraphics[width=6in]{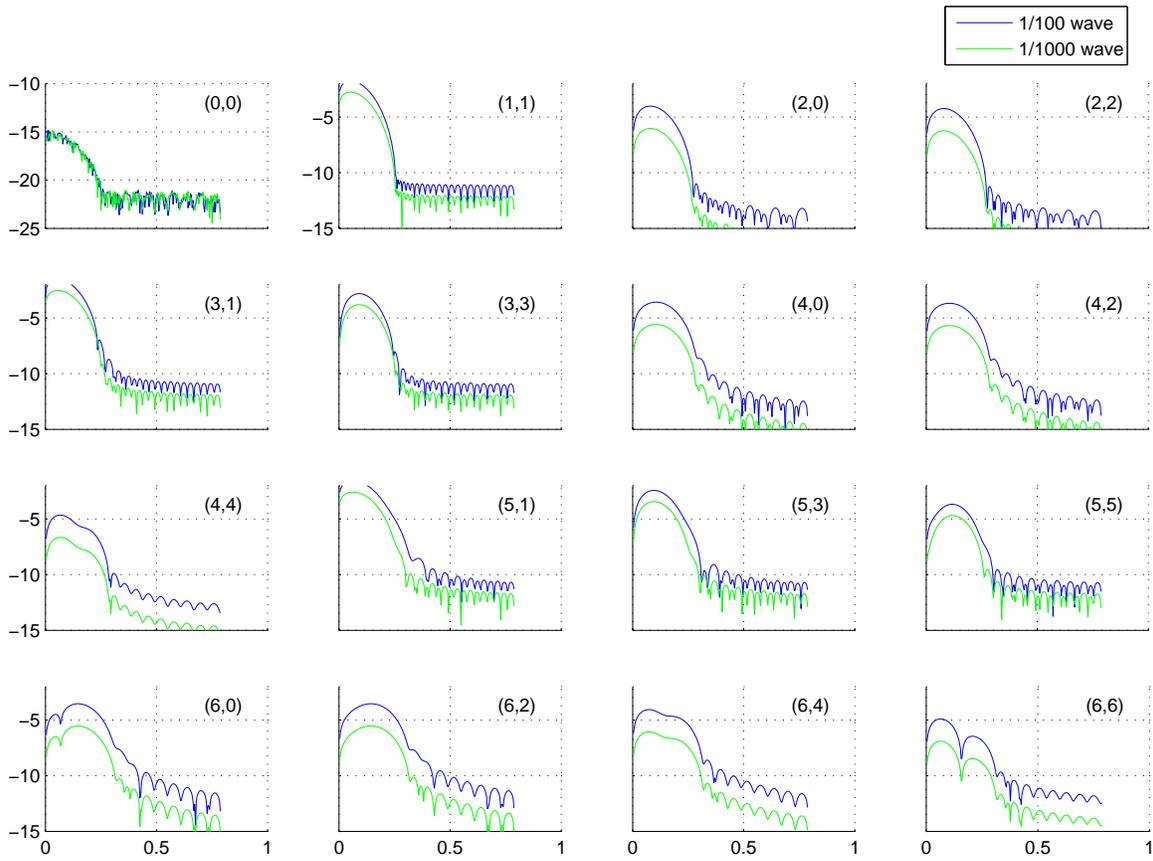}}
\end{center}
\caption{Difference plots.  Differences between the aberrated profile and the corresponding ideal.
}
\label{fig:10}
\end{figure}

In this paper we presented an efficient method for calculating the
distortions in the PSF of a pupil mapping system due to wavefront
aberrations. Figures \ref{fig:3} to \ref{fig:10} show that our
particular pupil mapping system is somewhat more sensitive to low
order aberrations than the concentric ring masks.  That is, contrast
and IWA degrade more rapidly with increasing rms level of the
aberrations.  However, this is partially mitigated by the
magnification property of pupil mapping, making a direct comparison
of the two systems more subtle, especially considering the
distortion in pupil mapping. (A more direct comparison can be made
for a pupil mapping that undoes the distortion, such as the
unmapping system described at the end of Section \ref{sec:hybrid},
but that is outside the scope of this article.) In any case, it is
evident that especially fine control of static and dynamic stability
will be required in order to take full advantage of the smaller IWA
intrinsic to the pupil mapping approach. A careful stability
analysis of any design employing pupil mapping is thus necessary to
determine its achievable operating range.

Finally, we note that there is a spectrum of apodized pupil mapping
systems. The two extremes, pure apodization and pure pupil mapping,
both have serious drawbacks.  On the one end, pure apodization loses
almost an order of magnitude in throughput and suffers from an
unpleasantly large IWA.  At the other extreme, pure pupil mapping
fails to achieve the required high contrast. There are several
points along this spectrum that are superior to the end points.  We
have focused on just one such point, which was suggested by
\cite{GPGMRW05}. We leave it to future work to determine if this is
the best design point. With this paper, we have provided the tools
to analyze the sensitivity of these kinds of designs.

{\bf Acknowledgements.}
This research was partially performed for the
Jet Propulsion Laboratory, California Institute of Technology,
sponsored by the National Aeronautics and Space Administration as part of
the TPF architecture studies and also under JPL subcontract number 1260535.
The third author also received support from the ONR (N00014-05-1-0206).

\bibliography{}
\bibliographystyle{plainnat}   


\clearpage

\end{document}